# Improvement of Converter Surface Plasma Sources

V. Dudnikov[1,a)]

[1] Muons Inc. 552 N. Batavia Ave. Batavia, IL 60510. USA

[a)] Corresponding author: dvg43@yahoo.com

**Abstract**. A large volume surface plasma source (SPS) with a biased converter was developed for the Los Alamos linear accelerator. A large gas-discharge chamber with a multipole magnetic wall and 2 heated cathodes can support a discharge generating plasma. A cooled converter with a diameter of 5 cm and a potential of up to -300 V bombarded by positive ions and emits secondary negative ions, accelerates them and focuses in an emission aperture with a diameter of 6.4 mm. From this SPS, up to 18 mA of H- ions are extracted at a duty cycle of up to 10%.
The emitted H- beam current is attenuated through H- destruction in thick layer of gas and discharge plasma. The H- beam intensity and H- generation efficiency can be increased by decreasing the gas and plasma layer thickness between the converter surface and the emission aperture. It is possible to improve beam characteristics by making a small modification to the converted SPS. We propose to use a thin Penning discharge in front of the converter. The magnetic field for the Penning discharge is created by permanent magnets. Decreasing the plasma end gas layer between the converter and the emission aperture can decrease the H- beam loss and thus increase the extracted beam intensity up to 2 times.

## INTRODUCTION

The discovery of the cesiation effect produced a significant enhancement in negative ion emission from discharge after adding small amount of cesium, or other substances with low ionization potential [1] open the door for production high intense high brightness beams of negative ions [2,3,4,5] .

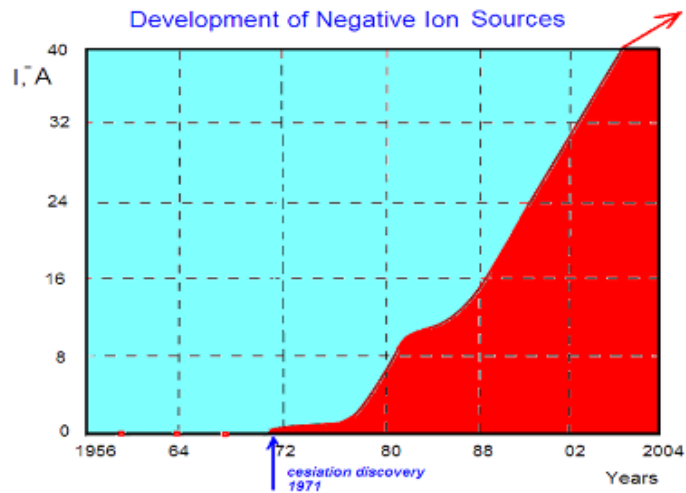

**FIGURE 1**. History of development of negative ion sources. Growth of beam intensity in time (Reproduced from V. G. Dudnikov, "Surface Plasma Method of Negative Ion Production", Dissertation for doctor of Fis-Mat. nauk, INP SBAS, Novosibirsk 1976).

Since the discovery of the cesiation effect the H⁻ beam intensity has increased to $10^4$ times from record 3 mA to > 40 as shown in Figure 1. Emission current density was increased from 6 mA/cm² to 6 A/cm² (1000 times).

The efficiency of beams negative ions generation was significantly increased by invention in of the geometric focusing of the generated negative ions [6,7], invented in Novosibirsk. The geometric focusing scheme is illustrated in Fig. 2. Secondary negative ions emitted by an emitter are accelerated in the near-electrode potential drop along a normal to the surface. If the surface is cylindrical or spherical, negative ions are focused to the center of curvature

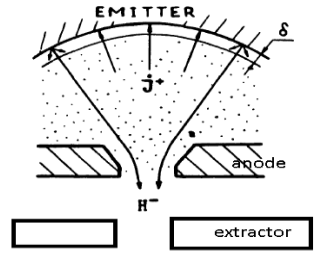

**FIGURE 2.** The geometric focusing scheme (Reproduced from V. G. Dudnikov, "Surface Plasma Method of Negative Ion Production", Dissertation for doctor of Fis-Mat. nauk, INP SBAS, Novosibirsk 1976).

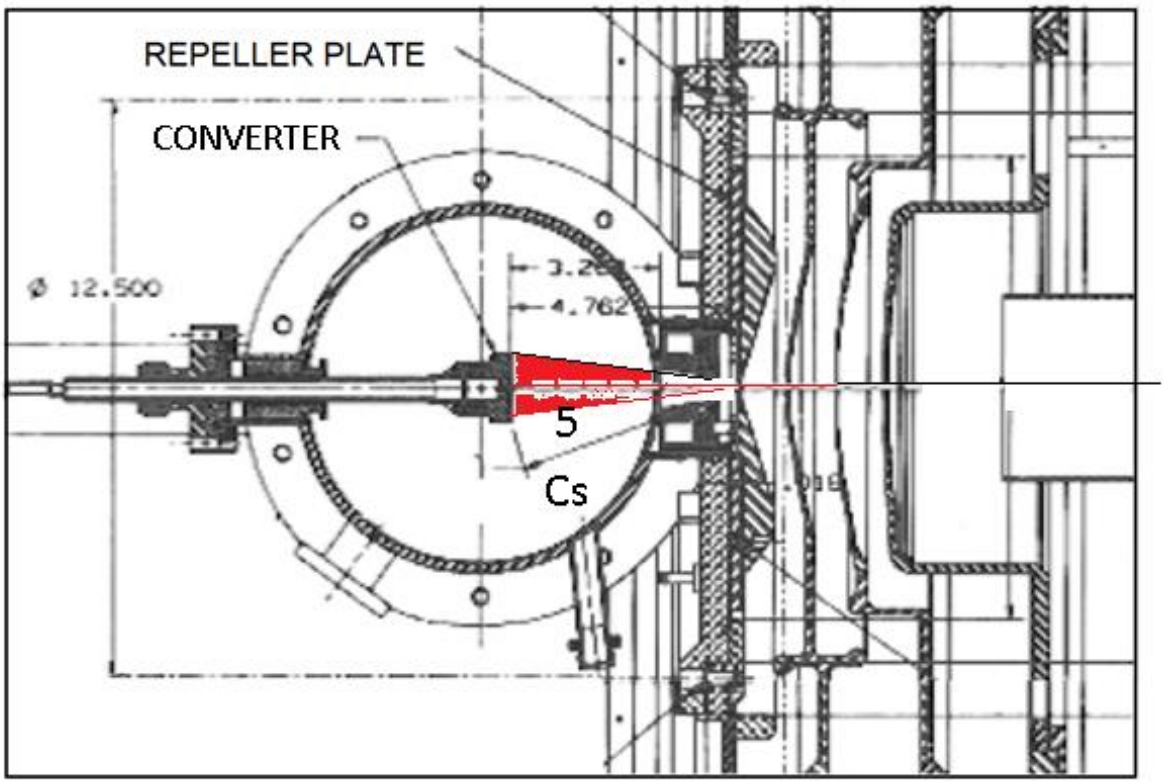

**FIGURE 3** (a) Schematic SPS with a converter for the Los Alamos Linear Accelerator. The beam emerging from the converter is indicated in red.

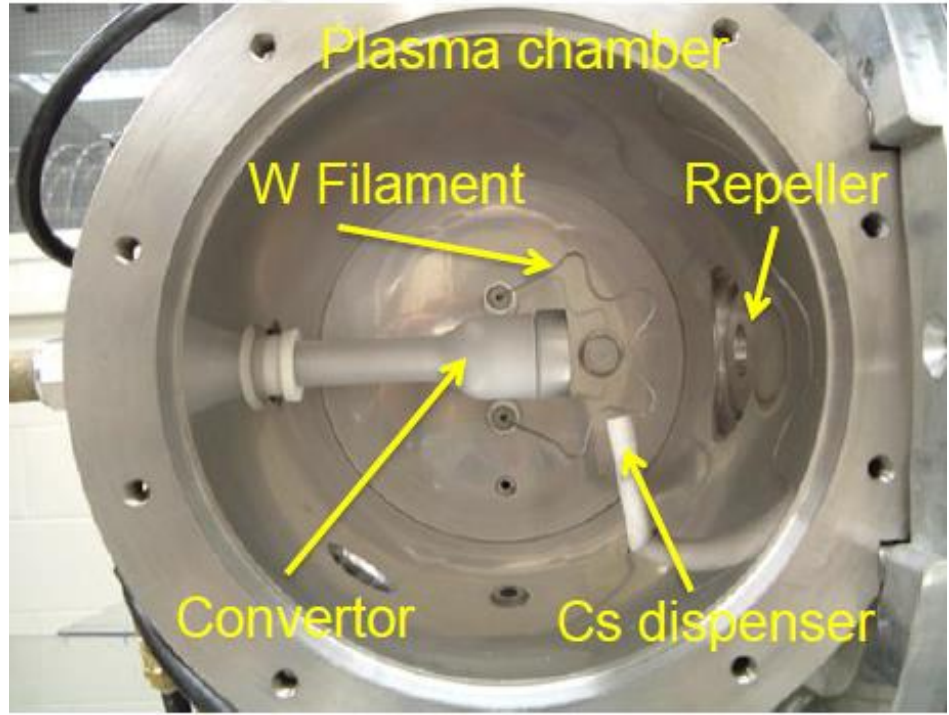

**FIGURE 3** (b) Photograph of converter SPS and associated elements (Reprinted with permission from Ilija N . Draganic, and Lawrence J. Rybarcyk, "Recent results in modeling of LANSCE H- surface converter ion source. AIP Conference Proceedings 2011, 050010 (2018)". Copyright 2018 American Institute of Physics).).

. On basis of cesiation effect a large volume surface plasma source with cesiation and geometrical focusing converter was developed in LBL [8]. On the basis of this LV SPS with converter, a smaller version of LV SPS with a converter for the Los Alamos linear accelerator was developed [9]. The schematic of this source is shown in Figure 3 (a), and photograph is shown in Fig. 3 (b). A smaller version of the LV SPS with a converter for the Los Alamos linear accelerator was also developed [9].

A large gas-discharge chamber with a multipole magnetic wall has a diameter of 17.8 cm and a height of 12.8 cm. Two heated cathodes with a diameter of 1.5 mm and a length of 20 cm support a discharge with a voltage of 90 V, generating a plasma with a density of up to $3 \times 10^{12}$ cm$^{-3}$. A cooled converter with a diameter of 5 cm and a potential of up to -300 V emits negative ions, accelerates them and focuses them in an emission aperture with a diameter of 6.4 mm. The distance from the emitter surface of the converter to the emission aperture is 12.5 cm (in the compact SPS this distance is < 1 mm). From this SPS, up to 18 mA of H$^-$ ions are extracted at a duty cycle of up to 10%. The normalized emittance of this beam is 0.13 π cm mrad.

For dependence of an extracted H- beam current on discharge power is typical a slow growing with increase the discharge power through H- destruction in thick layer of gas and discharge plasma between converter and emission aperture [10].

A six-filament version of the surface converter source with axial beam extraction was developed in a collaborative effort between Lawrence Berkeley National (LBNL) and Los Alamos National Laboratories (LANL) [11]. Although this source did produce 40-mA of H- beam current, an unexpected emittance growth factor of 2.5 made these higher current beams unacceptable for LANSCE operations [12]. A subsequent decision was made to fabricate and develop a fourth production source to produce 25-mA if current with an emittance growth of no greater than 20% [3]. The source upgrade technology would be used in a 750-keV H- injector B at LANSCE.

## IMPROVEMENTS TO CONVERTER SPS

The H- beam intensity, SPS lifetime and H- generation efficiency can be increased by decreasing the gas and plasma layer thickness between the converter surface and the emission aperture. It is possible to further improve beam characteristics by a small modification of the converter SPS as shown below. A schematic of this modification is shown in Fig. 4. It uses a thin Penning discharge in front of the converter, shifted to emission aperture. It can be used with the same discharge chamber with the hot filament located in front of anode with a slit for electrons to pass, and an anticathode for reflection of electrons. The magnetic field for the Penning discharge is created by the permanent magnets. A decrease of the plasma end gas thickness between the converter and the emission aperture can decrease the H- beam loss and increase the extracted beam intensity up to 2 times. The H- ion beam is extracted by the existing extraction system.  The gap between converter surface and emission aperture will be decrease up to 4-5 cm from 12.5 cm. Penning discharge have better confinement of fast electrons and can be started  at lower gas density then original converter source discharge. Estimation from Oak Ridge hot cathode Penning discharge give for gas pressure < 1 mTorr [13]

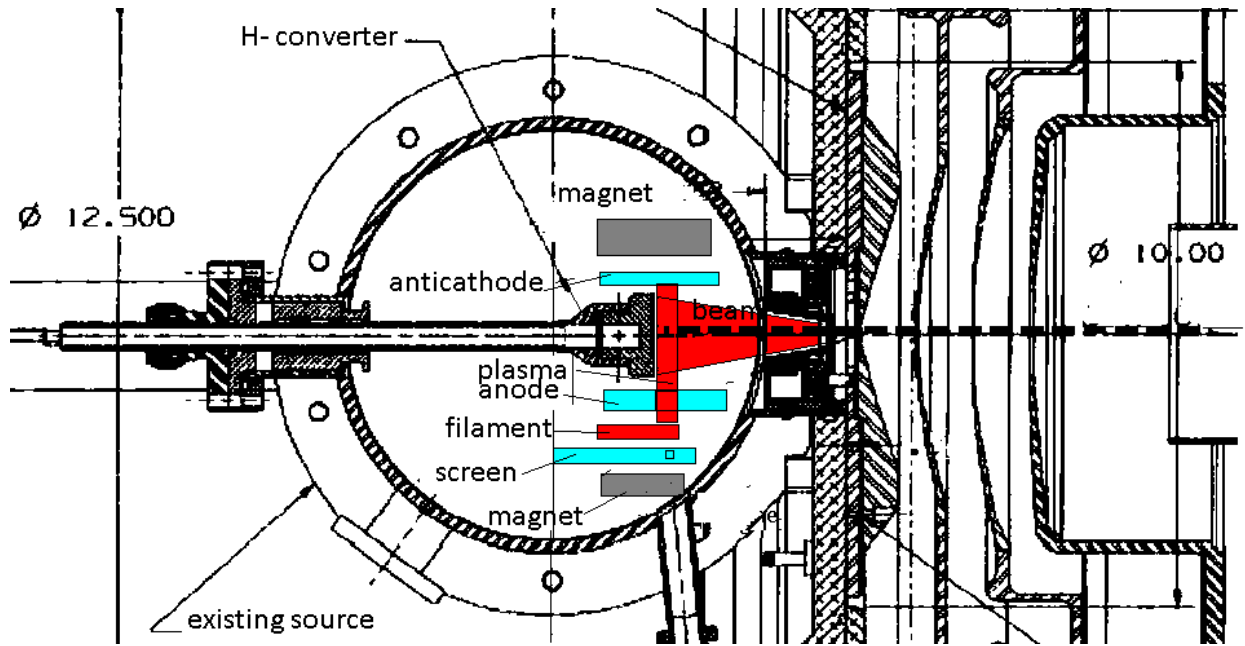

**FIGURE 4.** - Modified LANSCE converter SPS is shown. The heated cathode Penning discharge is shown below the converter and the anticathode shown above the converter elements.

The gas for penning discharge can be supplied by pulsed gas valve [14]. Fig. 5 is shows the heated tungsten cathode of the Penning discharge. The cathode and anticathode can be grounded. The discharge voltage can be applied to the anode. Magnetic field lines created by permanent magnets have a curvature which can be matched with surface curvature of converter. Plasma surface can touch the converter surface. It is don't necessary excellent matching of magnetic lines curvature with converter surface curvature. Magnetic field lines could cross the converter surface. In this case converter surface can service as anticathode reflecting of fast electrons and keep high ionization efficiency.

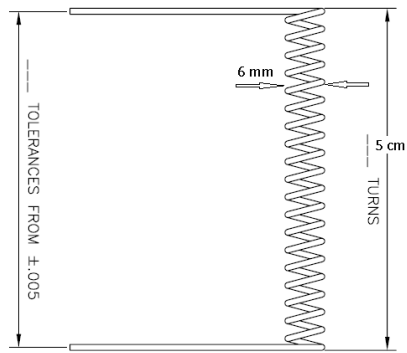

**FIGURE 5.** Heated tungsten cathode coil and dimensions of the Penning discharge.

Larmor radius of ions H- with energy W=400 eV in magnetic field 300 Gauss is

R=144 W$^{1/2}$/B=144 20/300~10 cm. In magnetic field area 1 cm H- ions can be bend for angle 0.1 rad. This bending can be compensated by repeller magnetic field. A little vertical shifting of the ion beam can be compensated by shifting of the conveter. A cathode as shown in Fig. 5 was used in [15] and show longer lifetime in DC mode of operation,. much longer then original TRIUMF cathodes similar to LANSCE cathode.

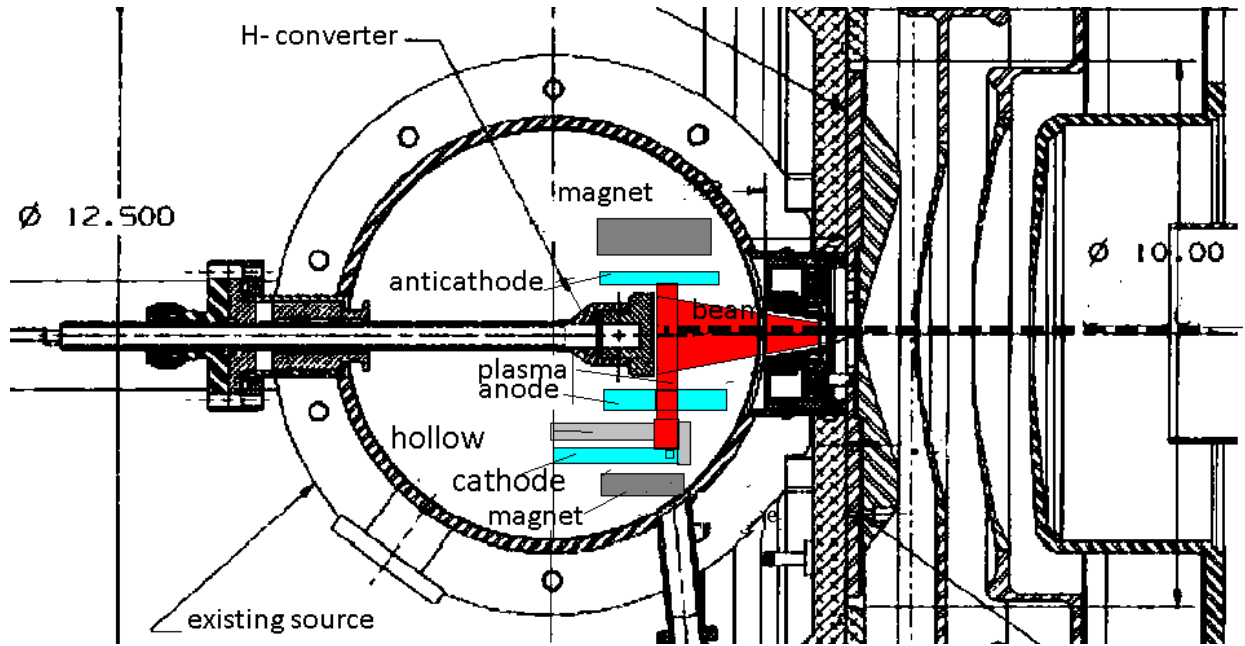

**FIGURE 6**. Modified LANSCE converter SPS with a hollow cathode Penning discharge elements.
.
Another possibility is to use a hollow cathode Penning discharge for generation of a thin plasma sheet as shown in Fig. 6. A thin layer of plasma near the converter surface is created by discharge in the magnetic field with a hollow cathode and anode. It was tested in the SPS with separated functions [4,5]. A schematic of an SPS with separated functions is shown in Fig. 7

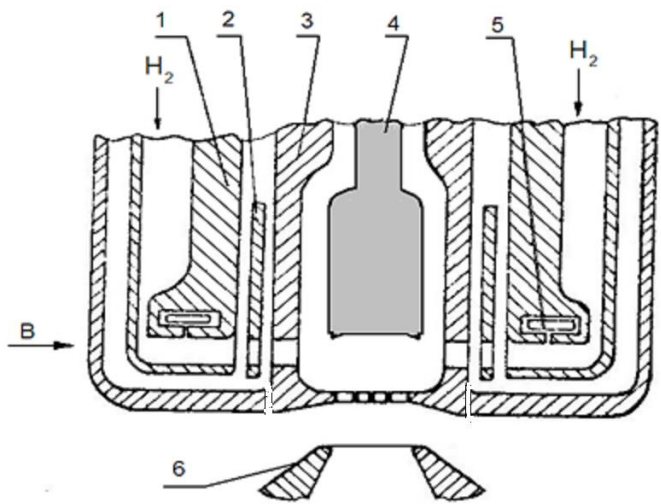

**FIGURE 7**. Schematic of an SPS with separated functions [4]. The elements shown are: 1– cathodes of the plasma generation system, 2 – bounding diaphragms, 3 – anode, 4 – independent emitter, 5 – cavities for cesium, 6 – extraction electrode (Reproduced from V. G. Dudnikov, "Surface Plasma Method of Negative Ion Production", Dissertation for doctor of Fis-Mat. nauk, INP SBAS, Novosibirsk 1976).

Experiments to optimize the processes considered were attempted in an SPS with separated electrode functions. In these sources, positive ions were generated by an independent discharge with a separate electrode system and a special electrode was used as the H⁻ ion emitter. The energy of the bombarding particles extracted from the plasma into the emitter was regulated by changing the potential difference between this electrode and the gas discharge plasma.

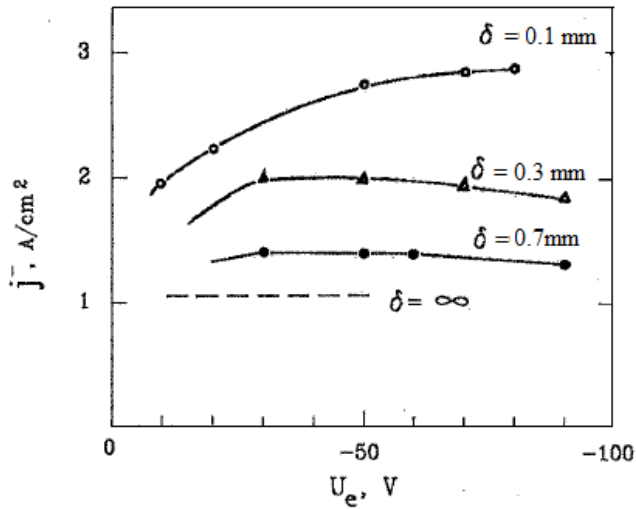

**FIGURE 8.** Performance of a separated function SPS. Current density of H⁻ions as a function of the voltage the emitter is shown for several different positions of the emitter relative to the plasma [ (Reproduced from V. G. Dudnikov, "Surface Plasma Method of Negative Ion Production", Dissertation for doctor of Fis-Mat. nauk, INP SBAS, Novosibirsk 1976).

Positive ions are formed in a flat layer of a gas-discharge plasma with transverse dimensions $0.12 \times 3$ cm$^2$ formed by establishing a discharge with electrons reflecting in the magnetic field between hollow cathodes with cavities in the form of long narrow slits and coaxial anode windows 3. The distance between cathodes in this SPS is 1.2 cm. Hydrogen is supplied by a pulse valve in the cathode cavity. To increase the efficiency of electron emission in the cathode cavity, cesium was released from cesium chromate tablets with titanium in cavities 5, when heated. The discharge voltage can be regulated over a wide range by changing the cesium release. Stable operation of pulsed discharges is obtained up to a discharge current of 450 A

The emitter 4 is installed with an adjustable gap Δ between its working surface and the boundary of the gas-discharge plasma layer (negative values of Δ correspond to its penetration into the plasma). Cesium for the emitter is fed from a container with independent heating. An emission hole with dimensions up to 0.3 x 3 cm$^2$ is located opposite the emitter working surface. To reduce the flow of accompanying electrons, the emission slit along the magnetic field is divided into three separate narrow slits by a jalousie (a shutter with a row of slats). with transverse dimensions 0.1 x 0.5 mm$^2$. The gas-discharge chamber is biased negatively at the extraction voltage.

## CONCLUSION

The proposed modifications of a converter SPS with cesiation and Penning discharge can improve negative ion production and increase the converter SPS lifetime.